# Synthetic aperture microwave imaging with active probing for fusion plasma diagnostics

**Vladimir F. Shevchenko,**[a] **Roddy G. L. Vann,**[b] **Simon J. Freethy,**[a,b] **Billy K. Huang**[b,c]

[a] *EURATOM/CCFE Fusion Association,*
   *Culham, Abingdon, Oxon, OX14 3DB, U.K.*
[b] *York Plasma Institute,*
   *Department of Physics, University of York, York YO10 5DD, U.K.*
[c] *CfAI,*
   *Department of Physics, Durham University, Durham, DH1 3LE, U.K.*
   *E-mail*: Vladimir.Shevchenko@ccfe.ac.uk

ABSTRACT: A Synthetic Aperture Microwave Imaging (SAMI) system has been designed and built to obtain 2-D images at several frequencies from fusion plasmas. SAMI uses a phased array of linearly polarised antennas. The array configuration has been optimised to achieve maximum synthetic aperture beam efficiency. The signals received by antennas are down-converted to the intermediate frequency range and then recorded in a full vector form. Full vector signals allow beam focusing and image reconstruction in both real time and a post processing mode. SAMI can scan over 16 preprogrammed frequencies in the range of 10-35GHz with a switching time of 300ns. The system operates in 2 different modes simultaneously: both a 'passive' imaging of plasma emission and also an 'active' imaging of the back-scattered signal of the radiation launched by one of the antennas from the same array. This second mode is similar to so-called Doppler backscattering (DBS) reflectometry with 2-D resolution of the propagation velocity of turbulent structures. Both modes of operation show good performance in a real fusion plasma experiments on Mega Amp Spherical Tokamak (MAST). We have obtained the first ever 2-D images of BXO mode conversion windows. With active probing, the first ever turbulence velocity maps have been obtained. In this article we present an overview of the diagnostic and discuss recent results.

# Contents

# 1. Introduction

Electron temperatures of several keV are typically achieved in present day magnetically-confined fusion experiments. Plasma, like any other matter, emits thermal radiation. In a magnetic field this emission is concentrated near the electron cyclotron (EC) frequency $\omega_{ce}$ and its harmonics due to electron gyro-motion around magnetic field lines. Thermal EC emission (ECE) in rarefied plasmas, where the plasma frequency $\omega_{pe}$ is much less than $\omega_{ce}$, can propagate through and escape the plasma in the so-called ordinary (O) and extra-ordinary (X) modes. Thermal ECE escaping the plasma carries information about its temperature and other important parameters. It has been used for diagnostic purposes in plasma experiments for decades.

In medium density experiments where $\omega_{pe} \geq \omega_{ce}$, ECE from the fundamental EC resonance may be partially blocked by plasma cut-offs. In this case, higher EC harmonics can be used for diagnostic purposes. In spatially-varying magnetic fields, complicated plasma geometries and in the presence of plasma turbulence, the information carried by ECE becomes multi-dimensional. Recent advances in microwave and millimetre wave technology have made it possible to develop and deploy ECE imaging systems in a number of fusion experiments [1]. These systems are based on optical approaches, i.e. they utilise large focusing optics to form an image on an array of detectors. An excellent review of imaging systems based on the optical approach can be found in [2].

In spherical tokamaks such as the Mega Amp Spherical Tokamak (MAST) [3] the plasma is usually well over-dense $\omega_{pe} \gg \omega_{ce}$. Thus fundamental and lower EC harmonics are completely confined by plasma cut-offs so that O and X modes do not propagate. However electrons still generate thermal emission at harmonics of the EC frequency. This emission is carried by electron Bernstein waves (EBWs) in over-dense plasmas. These electrostatic plasma waves cannot propagate outside the plasma but they can escape the plasma via mode conversion into electromagnetic waves. The mode-converted EBW emission is concentrated within narrow angular cones determined by the magnetic pitch angle and density gradient at the layer where the wave frequency coincides with the plasma frequency $\omega = \omega_{pe}$. EBW emission measurements over a range of viewing angles provide the angular positions of mode conversion cones, from which the magnetic pitch angle in the mode conversion layer can be found. Magnetic pitch measurements are of extreme importance for understanding plasma physics in tokamak and this sort of measurement has been demonstrated experimentally [4].

Motivated by these results, a Synthetic Aperture Microwave Imaging (SAMI) system has been designed and built to obtain 2D images of the mode conversion cones. The reasons for choosing synthetic aperture techniques instead of an optical approach become obvious if one looks at the frequency range in which emission must be imaged. In MAST plasmas the EBW emission frequency from the fundamental EC resonance covers the range of 10-18 GHz which corresponds to wavelengths from 30 mm to 17 mm. The optical aperture required for focusing of the EBW emission images at these frequencies would exceed any available vacuum window on the MAST vessel. In contrast aperture synthesis does not require any optics and a vacuum window of 150 mm diameter can provide satisfactory resolution of 2D structures in thermal EBW emission. In addition, active plasma probing with monochromatic waves can be easily adapted to SAMI, thereby providing 2D plasma turbulence velocity measurements simultaneously with passive EBE imaging.

## 2. Principle of synthetic aperture imaging

The synthetic aperture imaging was originally developed in radio astronomy [5]. The possibility of using aperture synthesis in a tokamak has been studied theoretically [6] but to our knowledge the work presented here is the first experimental implementation of this technique in a fusion experiment.

If we are able to record the phase and amplitude of the electric field at a number of different points, this allows us to use the principle of constructive interference to focus the imaging device by altering the phase of the recorded signals. If we digitally record the information in full vector form, we are able to perform this focusing after the fact, choosing from any number of focusing schemes. In the case of conventional imaging, the total number of constraints on the image is equal to the number of antennas $N_a$, as the image is made up of $N_a$ pixels. In the phase-sensitive case, however, the reconstructed image is the result of a convolution of the real image with the array beam pattern. This beam pattern, as we shall see later, is made up of Fourier components from each antenna pair. This gives us a total of $N_a(N_a-1)/2$ constraints on the image. As the number of antennas rises we have of order $N_a^2$ image constraints in the phase sensitive case and only $N_a$ constraints in the conventional case. Provided the array configuration is designed to minimise the redundancy in these antenna pairs and minimise the level of oscillations in the array beam pattern, this can be a great advantage.

The aperture synthesis technique is based on the van Cittert–Zernike theorem [7, 8] which describes the propagation of spatial coherence. This theorem states that the complex visibility of a remote incoherent source is equal to the Fourier transform of the mutual coherence function.

$$\Gamma_{i,j}(u,v) = \iint G(\eta,\xi)I(\eta,\xi)e^{ik(u\eta+v\xi)}d\eta d\xi, \qquad (1)$$

where $u$ and $v$ give the number of wavelengths between points $i$ and $j$ along Cartesian axes in the observation plane; $\xi = \sin\theta$ and $\eta = \cos\theta \cdot \sin\phi$ are direction coordinates of a point of the distant source; $G$ represents the antenna gain pattern; $I$ is the intensity of the source; and $\Gamma$ is the mutual coherence function of complex signals $S_1$ and $S_2$ measured at two points $i$ and $j$ in the plane of observation:

$$\Gamma_{i,j}(u,v) = \int S_i(t)S_j^*(t-\tau)dt. \qquad (2)$$

Here star means complex conjugation and $\tau$ is the time lag between measurements at points $i$ and $j$. In the special case of $\tau = 0$ the mutual coherence function is called the visibility function.

Equation (1) can be identified as a sample of the two-dimensional Fourier transform of the product of the brightness distribution and antenna gain pattern at a point in Fourier space $\{u, v\}$. Knowing this, one can construct an array of antennas and measure the cross-correlations for each pair of antenna signals, thereby providing sampling of the Fourier transform of the image in front of the array. With a sufficient number of antenna pairs, one may perform an inverse Fourier transform to obtain an approximation to the real source pattern. A detailed discussion on the principles and applications of synthetic aperture imaging can be found in [5].

## 3. Synthetic Aperture Microwave Imaging (SAMI) on MAST

### 3.1 Block diagram of the system

Structurally the system consists of 4 modules: a fast frequency switching local oscillator source; an RF down-conversion module; an active probing source and a data acquisition module. The RF down-conversion module is mounted on one of MAST's vacuum ports and connected to the antenna array with coaxial cables. The remaining components are assembled in a standard 19-inch rack which is located at about 5 metres away from the machine. In the following subsections each module is described in detail.

### 3.1.1 Local oscillator source

In order to provide radial resolution, the local oscillator (LO) frequency must be able to scan across the required frequency range. The LO source should have small variations of power within the frequency range, low noise and good phase stability. A bank of fast switching fixed frequency oscillators was designed to provide the required parameters.

A schematic of the LO source is shown in Fig. 1. Dielectric resonator oscillators (DRO) were chosen as fixed frequency sources. DRO series oscillators manufactured by MITEQ produce ultra clean signals with spurious modes below -80dB and typical phase noise from -95dBc/Hz to -80dBc/Hz @ 10 kHz for the frequencies within the range from 5GHz to 18GHz. The following sixteen frequencies (in GHz) were chosen for the LO source: 5.0, 5.5, 6.0, 6.5, 7.0, 7.5, 8.0, 8.5, 9.0, 9.5, 10.25, 11.25, 12.25, 13.25, 14.25, 15.25 and 17.25. A second harmonic down-converter is employed, so this frequency set provides detailed coverage of the 10 - 34.4GHz range: this extends up to the 3$^{rd}$ EC harmonic in the MAST tokamak.

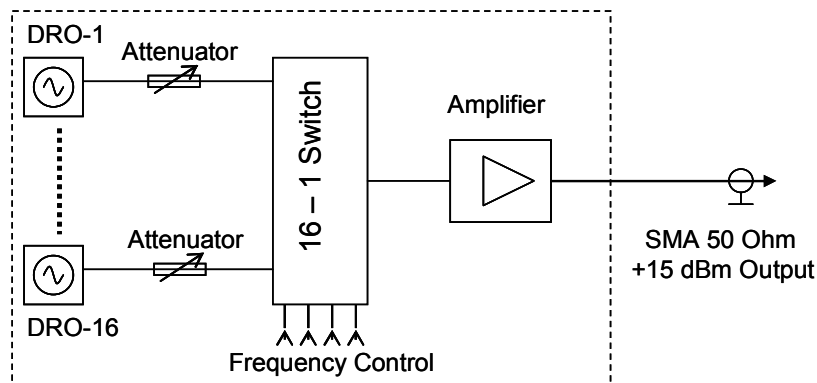

**Figure 1** Block diagram of the frequency switching local oscillator source.

All DROs are connected to the 16 throw 1 pole terminated absorptive switch via equalising attenuators. The switch supplied by PMI Inc. provides a switching time of 100ns with isolation better than 60dB and insertion loss of -7.5dB. The switch is controlled by TTL code either selected manually or generated by the acquisition & control unit described later. The chosen LO frequency signal is amplified by a power amplifier and delivered via 8m coaxial cable to the down-conversion unit (DCU) described in the next paragraph. In order to compensate different gain and transmission losses over the range of LO frequencies, equalising attenuators are employed to provide equal power at the mixers in the DCU. All components of the LO source are assembled in a standard 2U 19-inch rack unit.

### 3.1.2 Down-conversion unit (DCU)

A schematic of the DCU is shown in Fig. 2. The signal from the LO source is divided into four equal parts. Then the LO signal is amplified with power amplifiers ZVA-213+ supplied by Mini-Circuits. Each amplified signal is divided again into 4 parts which are equalised with attenuators to provide 16 identical +10 dBm outputs to feed the mixers. Equalisation of the LO signals at the mixer feed was conducted at the final stage when all the RF components and RF cables had been assembled. LO signals were balanced to +10±0.5 dBm across all channels and all frequencies.

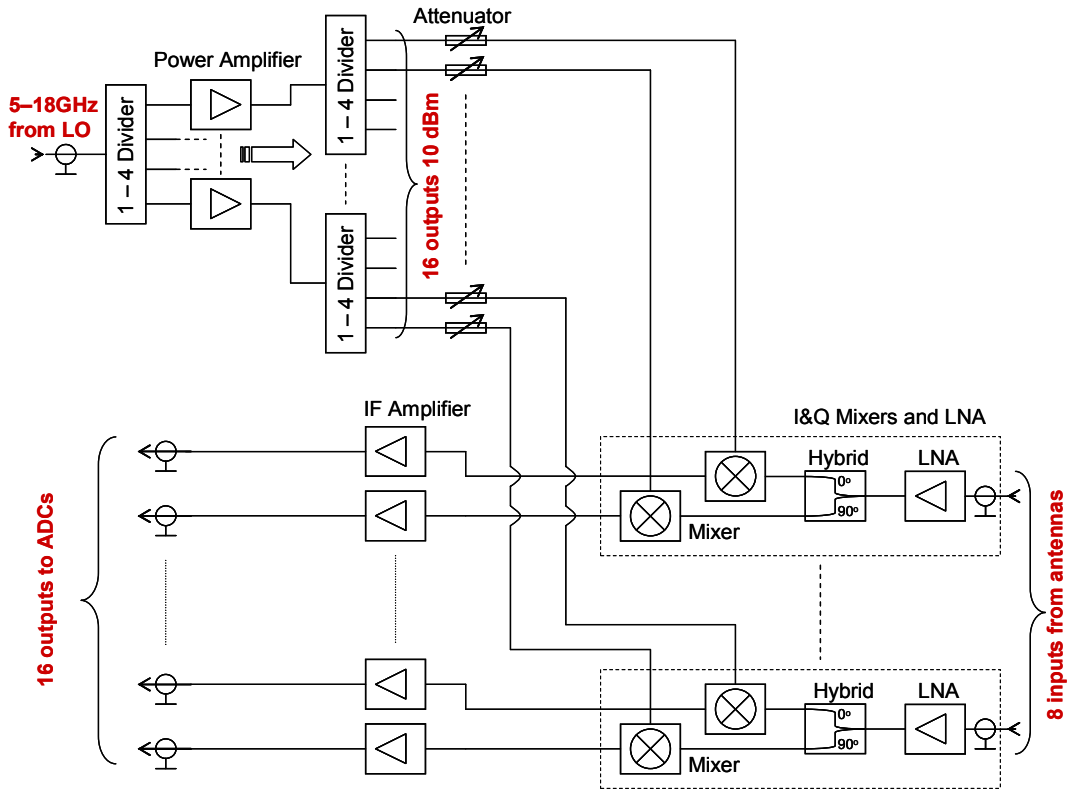

**Figure 2** Block diagram of the down-converter module.

Signals from antennas are amplified by 27 dB low noise amplifiers QLW-06404527 supplied by QUINSTAR. The RF signals are then split into 2 parts by 90 degree hybrid couplers. The output RF signals have $90^o$ phase-separated so-called in-phase (I) and quadrature (Q) components. These I and Q signals are input to the mixers. Second harmonic mixers SBE0440LW1 (MITEQ) are employed in the DCU allowing the use of LO frequencies a factor of two lower than the RF frequency. By using second harmonic mixers, all LO frequencies are kept below 18 GHz which makes it possible to transmit the LO signal through 8m coaxial cables from the LO source.

Intermediate frequency (IF) signals from the mixers span the frequency range from DC to 1.5 GHz. IF signals are amplified by 40 dB with ZKL-1R5 (Mini-Circuits) amplifiers which have a bandwidth from 10 MHz to 1500 MHz. These signals are transmitted then to the data acquisition module via 12 m coaxial cables. In order to protect the analogue-to-digital converters (ADCs) from over-voltage, the IF signals are passed though VLM-33+ limiters (Mini-Circuits) which keep the signal voltage within ±1 V. The video bandwidth of the ADCs is 700 MHz and their acquisition speed is 250 Mega samples per second. In order to avoid aliasing, the IF signals are passed though low pass filters SLP-100+ (Mini-Circuits) with pass band from DC to 98 MHz. Upper and lower side band separation is then performed numerically at the post-processing stage as described below in section 4.

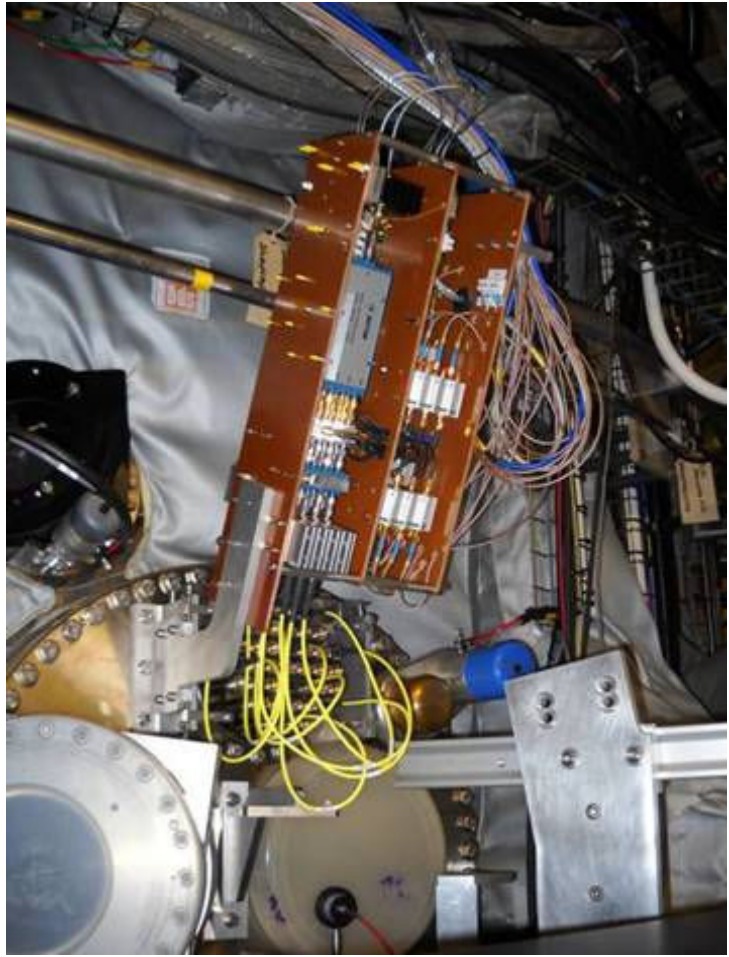

**Figure 3** Photograph of the RF down-conversion unit as installed on MAST with 8 antennas connected via high frequency (yellow) coaxial cables.

Coaxial-based technology is employed throughout the SAMI design which allows almost 4 octave frequency coverage using the same active components. All DCU components are assembled on an electrically-insulated frame attached directly to the vacuum vessel (see Fig. 3). The antennas are connected to the DCU via short coaxial cables which have extended operational range up to 40 GHz. They are bright yellow in colour and can be clearly seen in the photograph.

### 3.1.3 Data acquisition and control unit (DACU)

The data acquisition and control unit (DACU) is based on field programmable gate array (FPGA) technology. Two Xilinx ML605 FPGA boards are used to control two FMC108 ADC boards (4DSP). These boards, together with their power supply and optical fibre interface are incorporated into a 4U 19-inch rack unit as illustrated in Fig. 4. The DACU provides 16 ADC channels digitised at 250 Mega samples per second each with 14-bit resolution. Each ADC channel is equipped with fast onboard memory allowing continuous data acquisition for 0.5s. A detailed discussion of the DACU architecture and performance can be found in [9].

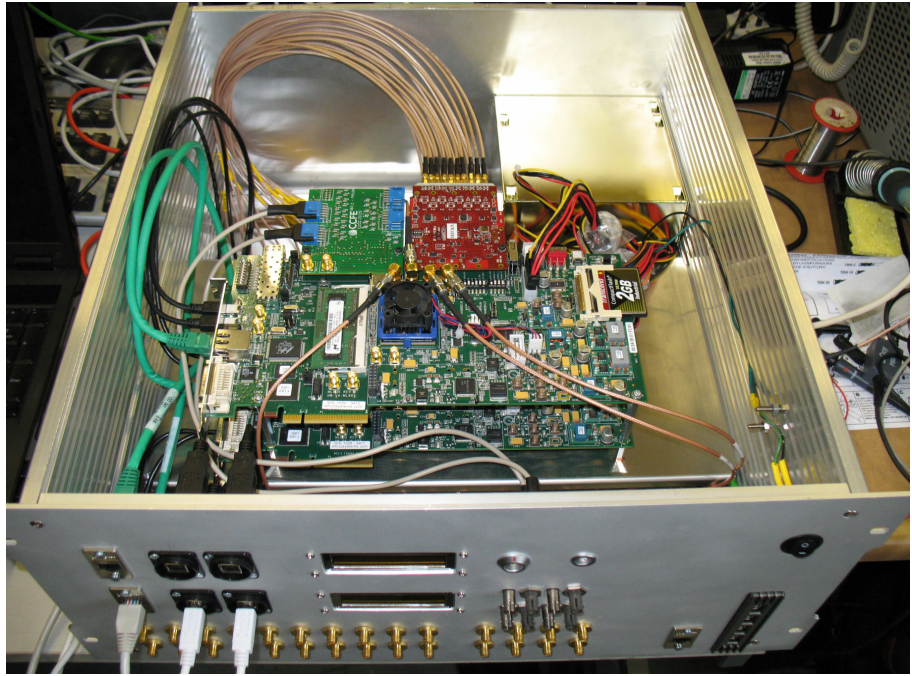

**Figure 4** FPGA-based data acquisition and control module assembled in a 4U 19" rack unit.

FPGAs also provide a number of control functions for the entire system. They generate clock frequencies of 10 MHz and 250 MHz synchronized either locally or with the central MAST experimental clock system. They generate a trigger signal and sequence of codes for LO frequency control. Configuration of the system can be reprogrammed between plasma shots. Any combination of LO frequencies can be chosen with a switching frequency from 1 to 1000 μs. The FPGAs also control the DCU power sources which thereby supply power to the amplifiers only during the plasma pulse. Pulsed power of the DCU allows achievement of high gain stability across all channels and antennas. FPGA boards support continuous communication between the DACU and remote computers via a fibre optic link. One of the boards generates a set of IF signals at 8, 10 and 12 MHz for the active probing source which is described in the next subsection.

### 3.1.4 Active probing signal generator (APSG)

The active probing signal generator (APSG) is built on the principle of a single sideband up-converter. Hybrids are used in an unusual manner due to the second harmonic mixers SBE0440LW1 employed in the up-converter as illustrated in Fig. 5. This technique utilises the same LO signal as the DCU and generates the probing signal at a frequency twice that of the LO, which is also up-shifted by the small IF frequency. By changing $90°$ hybrid orientations, either upper or lower sideband can be chosen at the output. The APSG can operate with any IF frequency within the range of the $90°$ hybrid. It can also work with a low frequency (10 – 100 MHz) noise source to provide a narrow-band RF noise signal for calibration purposes and noise probing experiments.

In the present DACU configuration, up to three APSGs can be used simultaneously in plasma experiments. IF frequencies are synchronized with the system clock and their values have been chosen to provide minimum interference between probing channels and minimize effects on passive imaging. In future, the low frequency noise source, IF signals and IF hybrid will be replaced with a single FPGA board generating either monochromatic or noise type I and Q components. This FPGA-based system will simplify the APSG schematic and improve side band suppression at the output.

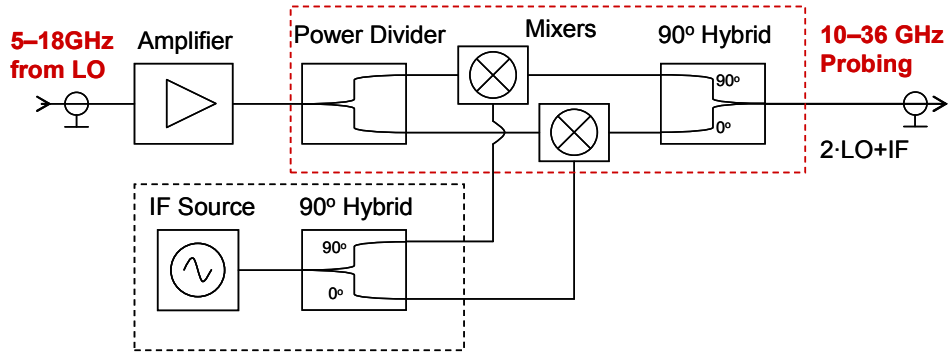

**Figure 5** Block diagram of the active probing source.

### 3.1.5 Antenna array design

The antenna array is one of the most important parts of the system. It must provide wide viewing angles both vertically and horizontally. It must provide extremely wide frequency coverage from 10 to 36 GHz with gain variations within 5-10 dBi. Such a performance with multi-octave frequency spans can be provided only by ridged horn antennas or so-called Vivaldi antennas [10]. A quad-ridged horn antenna would be ideal for microwave plasma imaging because it provides simultaneous measurements at two perpendicular polarisations. However their cost and physical size turned our preferences toward a Vivaldi type antenna.

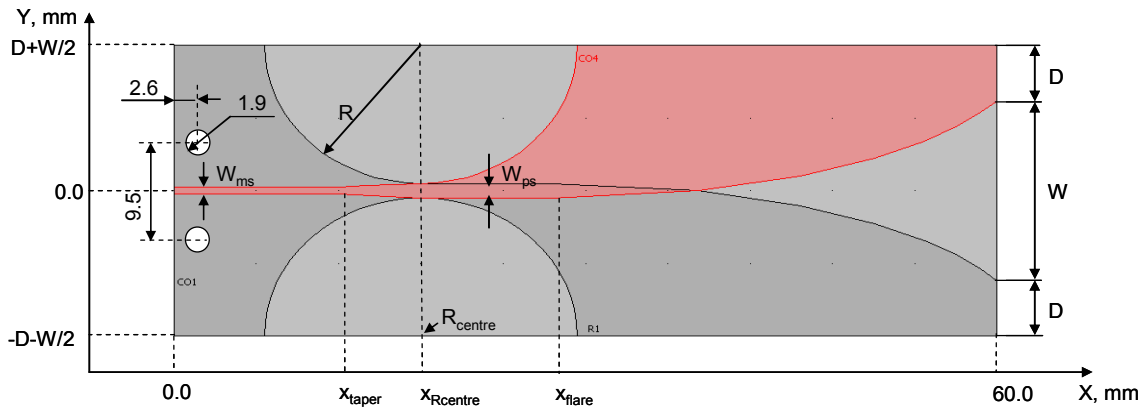

**Figure 6** Design of the antipodal Vivaldi antenna.

Vivaldi antennas can be made very small and accurately using printed circuit board (PCB) technology. An antipodal design proposed in [11] made it possible to achieve multi-octave frequency coverage. In our design we used Rogers RT5880 duroid with $\varepsilon_r = 2.2$, thickness of 0.381 mm, and 35 μm of copper metallization. A PCB design is illustrated in Fig. 6. Here the signal side copper is shown in red and the 'ground' side copper is shown as dark grey. The antenna is 60 mm long and 20 mm wide with a precision K-2.92 end launch jack (manufactured by Connector Gage Company, not shown) attached to the micro-strip line. The geometry of the antenna is described by the formula:

$$y = c_1 \cdot e^{T \cdot x} + c_2, \qquad (3)$$

with the flare opening parameters $T$ = 0.157 mm$^{-1}$, $c_1$ = 6.275·10$^{-4}$ mm and $c_2$ = -0.739 mm. The micro-strip line conductor width is $W_{ms}$ = 1.14 mm and the parallel strip-line conductor width is $W_{ps}$ = 1.46 mm. Remaining metallization is covered with gold to protect from oxidisation.

An antenna array assembled on a polycarbonate holder which contains 21 Vivaldi antennas is shown in Fig. 7. Individual antennas can be installed either vertically or horizontally providing sensitivity to any polarization. In practice, antenna spacing and relative positioning are limited by the available port window. The antenna arrangement shown here can be accommodated into the available port. At present, the antenna array is installed within a vacuum window of 150 mm diameter which is located 20 cm above the midplane of MAST. Only signals from 8 antennas are recorded by the DACU so the remaining antennas can be used for active probing.

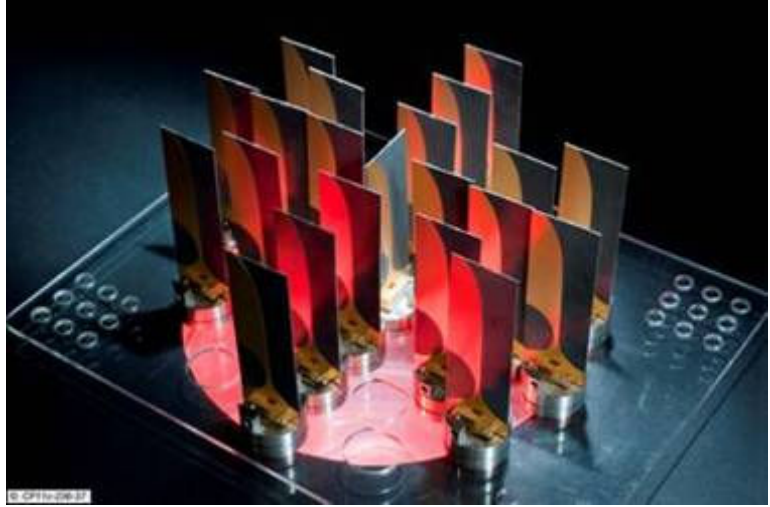

**Figure 7** A photograph of the antenna array with 21 antipodal Vivaldi antennas.

The number of antennas and their mutual positioning within the array are the primary factors defining image reconstruction quality. A special numerical method for array optimization has been developed [12] and was employed for our system. Several antenna arrays were designed using this method to obtain best imaging of the mode conversion windows using only 8 antennas or 16 antennas divided into 2 sets of 8 with fast switching between them.

### 3.2 Calibration of the system

An absolute calibration of the 2D synthetic aperture imaging system would require large 2D 'hot' and 'cold' black body sources. However for the correct image shape reconstruction a relative phase calibration is sufficient. We therefore performed thorough phase calibration using several techniques, but did not carry out an absolute calibration of the system.

The calibration was done on the machine in-situ with all components in the operational position except for the antennas. The antenna array was turned around and pointed to the outside of the machine as shown in Fig. 8. All cables were kept plugged-in and care was taken to disturb connectors as little as possible. The antenna array was surrounded by absorber sheets providing viewing angles about $\pm 20°$ horizontally and about $\pm 10°$ vertically. A dual-polarized horn antenna DP241-AC (Flann Microwave) was placed at 1.2 m from the array. The horn antenna can be moved both vertically and horizontally providing a 2D scan of the image plane with the active point source.

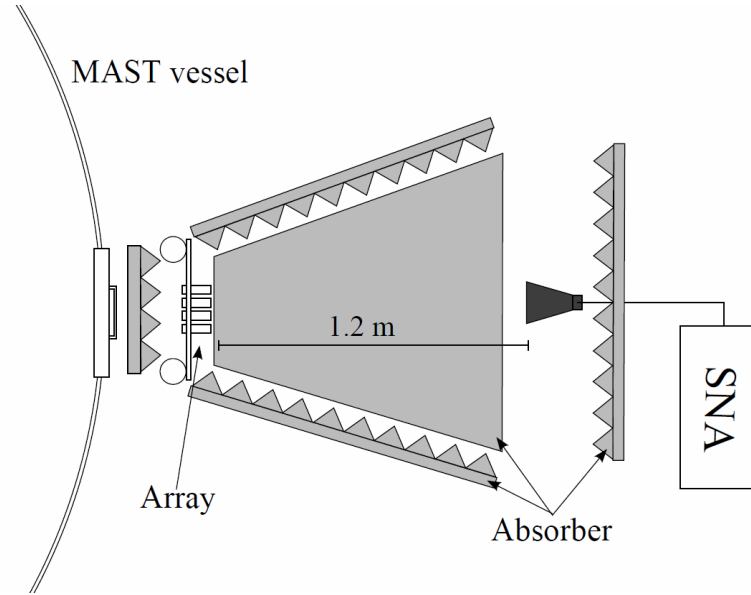

**Figure 8** Ex-vessel calibration set-up.

First a frequency sweeping (Rohde & Schwarz) microwave generator covering the range from 9.5 to 18 GHz was used to feed the horn antenna for calibration of low frequency channels. The antenna was moved in 2 cm steps providing detailed coverage of the viewing field. SAMI was run in a frequency-switching mode receiving all frequencies from 10 to 18 GHz. During the time instances when the frequency of the sweeping generator was passing the doubled frequency of one of the LOs, a beating frequency occurred in IF signals in all channels. To obtain calibration constants, only one calibration is necessary with the source located at known angular coordinates. The remaining measurements within the array viewing field were used to estimate the accuracy of the point source position reconstruction.

A similar procedure was conducted with the movable horn antenna fed by the APSG generating noise signals as described in section 3.1.4. In this configuration, only one side band is calibrated at any one time. To complete the calibration, a measurement with the other side band must also be made. The reconstructed coordinates of the point source show good agreement with actual coordinates across the field of view of the antenna array. Typical errors are within $\pm 2°$ and they do not show any systematic behavior: they appear random.

Noise signals generated by the APSG were also used for calibration at higher frequencies. The procedure was similar to that described above. Calibration at higher frequencies was also conducted using APSG fed by the frequency-sweeping microwave generator in order to double the frequency and produce frequency coverage up to 36 GHz. The reconstructed coordinates were within ±2° of the actual ones. However the image contains a higher fraction in side lobes because the wavelengths are roughly half those in the lower frequency band.

In the calibration procedures described above the majority of SAMI components were kept as they would be used in plasma experiments. However cables connecting the antenna array to DCU were disturbed and the vacuum window in front of the array was not included at all. To assess possible effects of the window and cable bending, a through-vessel calibration was also performed. There are 5 windows at the opposite side of the machine allowing illumination of the array using an active source. These windows are located near the midplane providing little variation of the resulting vertical viewing angle; however the resulting horizontal viewing angle varies from -35° to +15°.

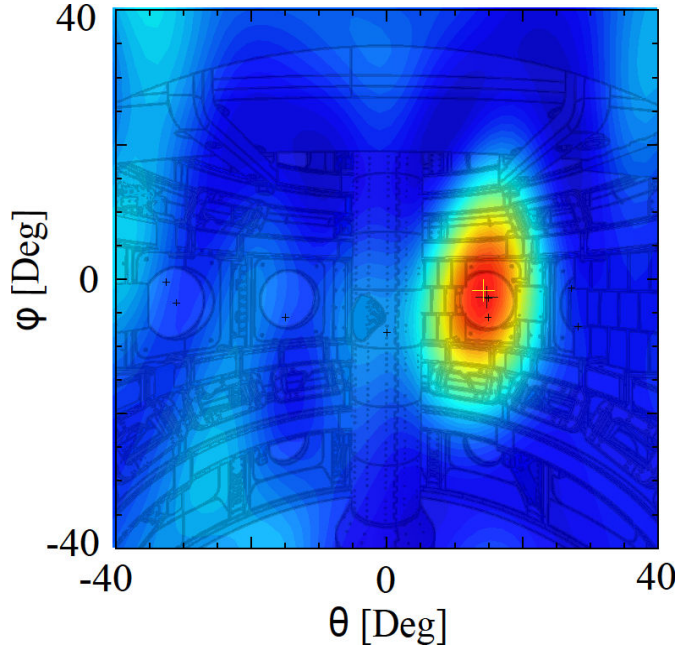

**Figure 9** Image of the noise source at 13 GHz shining through a port window in MAST Sector 2. Contours of the MAST vessel interior, as seen by the antenna array, are overlaid on top of the image.

Noise signals generated by the APSC were used in the through-vessel calibration. All available windows were used to illuminate the antenna array with the lower side band active and then measurements were repeated with the upper side band active. Fig. 9 represents the reconstructed image of the point noise source being shone through the vessel. Internal vessel components are over plotted. The large black '+' shows the actual position of the point source and the large white '+' indicates the maximum of the reconstructed image. The mismatch is less than 2° which is in agreement with off vessel calibration. The angular dimensions and elongated shape of the reconstructed image agree well with the shape of the antenna array used in this particular experiment. The array had an almost elliptical shape inclined by about 10° clockwise.

## 4. Data analysis

Traditionally in radio astronomy the data analysis started with a number of complex cross-correlation coefficients between all available pairs of antennas. These coefficients were usually generated at the hardware level. The RF signals from antennas were typically side band filtered and cross-correlated to each other in analogue multipliers and then the results were stored [5]. With recent advances in digital technologies a large fraction of the pre-processing functions can be moved onto FPGAs or computers [13, 14]. To our knowledge SAMI is the first system which acquires I and Q components of the IF signals in a digital form and the remaining analysis can be done numerically in real time or at the post processing stage.

### 4.1 Side band separation

SAMI records IF signals from each antenna in a full vector form (I&Q components). These signals can be represented in the form:

$$S_I = A_I[\cos(\delta\omega_u t) + \cos(\delta\omega_l t)]$$
$$S_Q = A_Q[\sin(\delta\omega_u t) - \sin(\delta\omega_l t)], \quad (4)$$

where $\delta\omega_u$ represents the range of frequencies within the upper side band and $\delta\omega_l$ represents frequencies within the lower side band. We used a Hilbert transform to obtain another pair of quadrature signals:

$$H(S_I) = A_I[\sin(\delta\omega_u t) + \sin(\delta\omega_l t)]$$
$$H(S_Q) = A_Q[-\cos(\delta\omega_u t) + \cos(\delta\omega_l t)]. \quad (5)$$

Obviously the side bands can be separated easily by simple algebra between appropriate pairs of signals from (4) and (5) assuming amplitude responses $A_I$ and $A_Q$ are equal or known from calibration. In practice side band separation is always limited by phase imbalance in the 90° hybrids. This imbalance can be partially corrected leading to improved side band separation. With appropriate phase correction, side band suppression better than -28 dB has been achieved in the whole range of SAMI frequencies.

### 4.2 Beam forming technique

There are a number of methods available for the analysis of synthetic aperture data. Most of them were developed for radio astronomy [5]. We used several of these methods in our image reconstructions, in particular one described in [15]. However here we will present a method based on a beam forming technique because it is equally applicable for the active probing data (for which the aperture synthesis method is not straightforward). Usually beam forming is used as a real-time technique to control the direction of the beam launched by the phased array of antennas or received from one or several particular directions [16]. We use the same method in a post processing stage by synthesising a beam and then virtually steering it over viewing angles.

A synthesized beam signal $S^B$ can be represented as a sum of the signals from all the antennas with weighting coefficients and phases:

$$S^B(t,\theta,\phi) = \sum_{i=1}^{N} w_i \cdot S_i^A[t, \varphi_i(\theta,\phi)], \quad (6)$$

where $w_i$ is a weighting coefficient obtained in calibration, $\varphi_i$ is a phase shift calculated for each $i$-th antenna for particular azimuth $\phi$ and elevation $\theta$ angles; $N$ is the number of antennas; $S_i^A$ is either an upper or lower complex side band signal from the $i$-th antenna. Due to the interference between signals from different antennas, the signal originating from the chosen viewing direction is preserved while signals from other directions are suppressed. In fact beam forming is a spatial filter searching for the photons originating in one chosen point in space and captured by all $N$ antennas. Then the synthesized beam can be steered in any direction within the angular range of the antennas by changing the phase shift of the complex signals $S_i^A$.

Moreover from the above one can see that the synthesized beam can be focused at any particular distance from the antenna array and then angular scanning can be done with the beam focused at a particular 3D surface, for example toroidal or spherical, which is the case for the present application. The beam focusing ability distinguishes the beam forming technique from aperture synthesis: aperture synthesis works only in the far field, i.e. it has the image plane placed at infinity. But beam forming works well in both the far field and the near field. This is important for experiments on MAST where the distance from the antenna array to the plasma surface is about 60 cm and the antenna array is fitted to a window of 15 cm diameter. A detailed discussion of near field effects and comparison of different methods used in image reconstruction will be the subject of a future publication.

### 4.3 Image reconstruction

The synthesized beam signal contains all the information necessary for passive image reconstruction and active probing data analysis. In order to build an image of thermal emission from the plasma, the active probing signals near 8, 10 or 12 MHz must be filtered out using a band reject or high pass filters. The remaining IF bandwidth from 15 to 100 MHz is used for passive imaging. The real part of the filtered signal $S^B$ is integrated over the exposure time interval $\Delta t$ and the result is then squared to generate one pixel in the intensity map.

$$I(\theta,\phi) = \{\int_{\Delta t} \mathrm{Re}[S^B(t,\theta,\phi)]dt\}^2 \qquad (7)$$

By scanning over a range of azimuth and elevation angles the whole 2D map of intensities is built. It must be noted that the number of 'meaningful pixels' is defined by the number of independent base lines $(N-1)\cdot N/2$. Thus the intensity map should have a larger number of pixels to provide a smooth image of thermal emission. Typical grid we use in image reconstruction is 20 by 20 while the number of independent baselines is 28.

In order to analyse the active probing signal, the synthesized beam signal $S^B(t)$ is Fourier transformed and the power spectrum is built. The spectrum around the probing frequency contains an un-shifted frequency component corresponding to reflection from the plasma and the vacuum window. That fraction of the spectrum is filtered out before further analysis. The spectrum also contains a frequency-shifted fraction corresponding to the signals which are back-scattered on fast-moving plasma fluctuations. We calculate the average frequency shift of this fraction and assign this value to the pixel at the corresponding viewing angles $\phi$ and $\theta$. This frequency shift is caused by the Doppler effect and so the corresponding 2D image represents a velocity map of the plasma turbulence. A complementary back-scattered power map can be plotted in the same manner as for emission imaging which may help in interpretation of the turbulence velocity map.

In fact both the thermal emission intensity 2D map and the turbulence velocity 2D map can be extended into a third spatial dimension by plotting a sequence of maps in a range of RF frequencies. For each RF frequency, a particular radial position can be assigned within the plasma and so complementary 3D maps can be constructed assuming that the background plasma and/or plasma flows did not change during the frequency switching cycle.

## 5. System performance and first results

The SAMI system performs in an automatic mode during experiments on MAST. It is synchronised with the central 10 MHz clock and trigger signals. SAMI performs active probing and image recording simultaneously over the range of pre-programmed frequencies for 0.5 second and then transfers the data to the data storage computer via a fibre optic link. During each 0.5s plasma shot, 4 GB of data are stored. During the 2011-2012 experimental campaign, the first ever 2D images of thermal emission from over-dense plasmas have been obtained and first ever 2D turbulence velocity maps have been recorded. These results are presented below.

### 5.1 Experimental results with passive imaging

Experiments were conducted on the MAST spherical tokamak. Plasma thermal emission was received within the frequency range from 10 to 35.5 GHz which cover the first three EC harmonics. The antenna array was attached to the port located 20 cm above the equatorial plane of the machine. The vacuum window was made of fused silica 15 cm in diameter.

Typically the plasma is significantly over dense in MAST – even during so-called low confinement mode (L-mode). In such a plasma the lower EC harmonics are obscured by cut-offs and thermal emission is mainly attributed to the mode-converted EBWs. First experimental images showed a presence of two distinctive bright spots located around [-20°, -20°] and [+20°, +5°] as illustrated in Fig. 10b. These angular coordinates correspond approximately to locations of mode conversion maxima for the B-X-O MC mechanism.

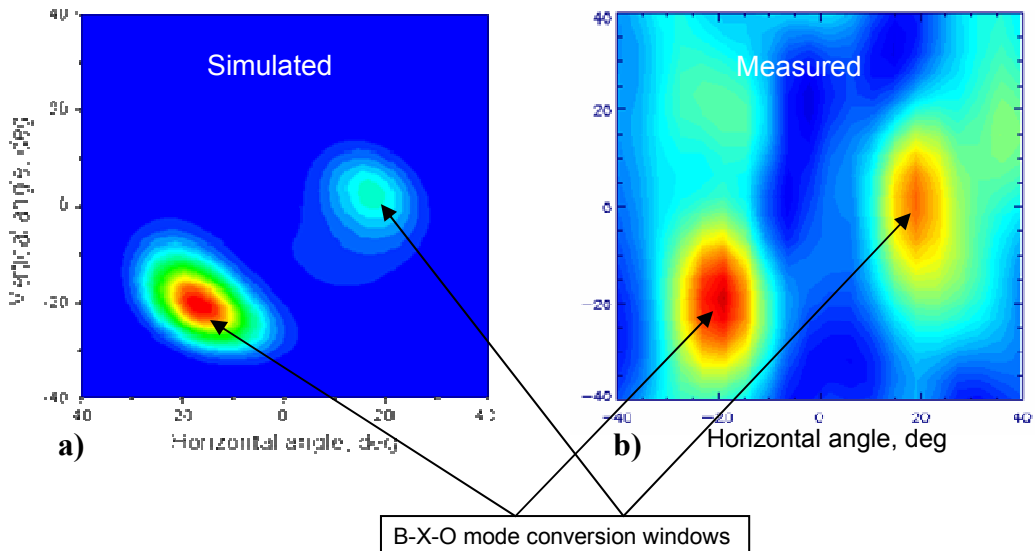

**Figure 10** Simulated (a) and measured (b) EBW emission from the fundamental EC resonance at 15 GHz in a low density L-mode plasma. Coordinates of mode conversion windows show excellent agreement. The vertical alongation of the measured image is consistent with the array convolution.

1D full-wave mode-coupling code complemented with 3D ray-tracing modeling [17] has been conducted for the same plasma parameters for comparison. Results are shown in Fig. 10a. One can see that the angular coordinates show excellent agreement with those experimentally observed. The experimental image has a well-pronounced vertical elongation, which is consistent with the shape of the antenna array used in this experiment. The other less obvious features of the image are likely to be caused by reflections within the vacuum vessel. Thus two main maxima were identified as mode converted EBW emission. To our knowledge this is the first ever experimentally obtained 2D image of the mode conversion windows in a fusion plasma.

### 5.2 Ex-vessel tests with active probing

SAMI has the capability of probing the plasma by monochromatic waves with further recording of the back-scattered waves. To test this potential an ex-vessel experiment has been conducted. A rotating assembly (see Fig. 11a) made of perforated aluminium forms several corner reflectors which return the incident emission back in the plane of rotation. This assembly was placed at about 1 m from the antenna array and it was rotated slowly during image recording. The same assembly of absorbers was used as for the ex-vessel calibration (see Fig. 8). One of the antennas within the array was used to probe the object with the APSG signal with 10 MHz up-shift as described in section 3.1.4.

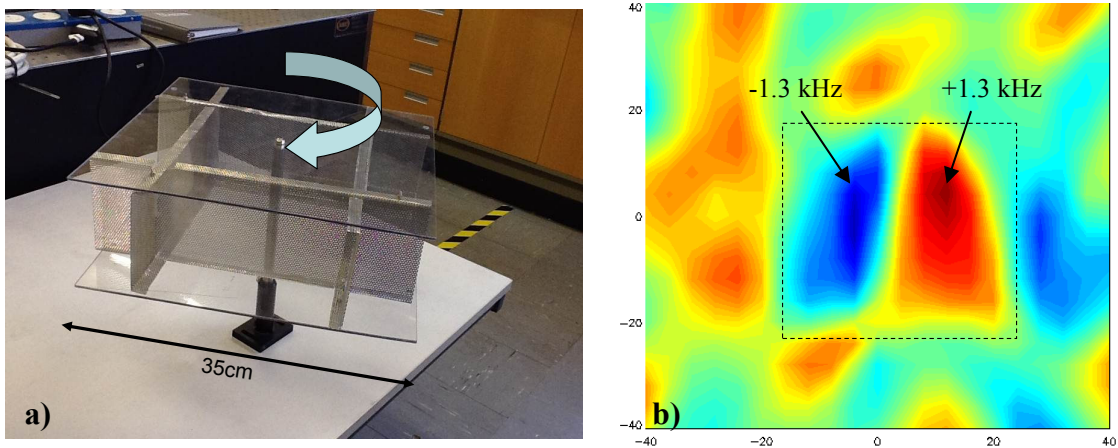

**Figure 11 a)** Rotating corner reflector was used in active probing ex-vessel tests. **b)** Velocity map reconstructed at 17 GHz. Colour represents frequencies: blue corresponds to negatively shifted; red to positively shifted frequencies. Maximum Doppler shifts of ±1.3 kHz are consistent with the rotation speed of the reflecting assembly.

The image recorded in this experiment at 17 GHz is shown in Fig. 11b. The area occupied by the corner reflector is indicated by the dashed line rectangle. Maxima of the frequency shifts within blue and red areas have the same magnitude of 1.3 kHz but have opposite signs. The measured frequency shift value is consistent with the Doppler shift due to the reflector rotation. This value is proportionally decreasing with the frequency decrease and it is increasing with the frequency increase. Thus it was concluded that the observed blue and red areas on the map are indeed caused by the Doppler shift of the reflected emission due to rotation.

A complementary intensity map shows that the main power is concentrated within the dashed rectangle while the spots outside the rectangle are probably caused by parasitic reflections from the absorbing mats.

### 5.3 Experimental results with active probing

As was already mentioned, SAMI can perform active probing of the plasma simultaneously with passive thermal emission imaging in order to produce 2D images of the signals back-scattered from the plasma. The majority of plasma experiments were conducted with 10 MHz probing because it is located right at the lower end of the IF range and it gives minimum disturbance to the imaging bandwidth. A non-coherent back-scattered signal is received within the upper side band only. The reflected coherent fraction without Doppler shift is filtered out. Similarly to the passive imaging technique, the antenna phasing steers the viewing direction and generates a 2D image of Doppler shifts for every frequency.

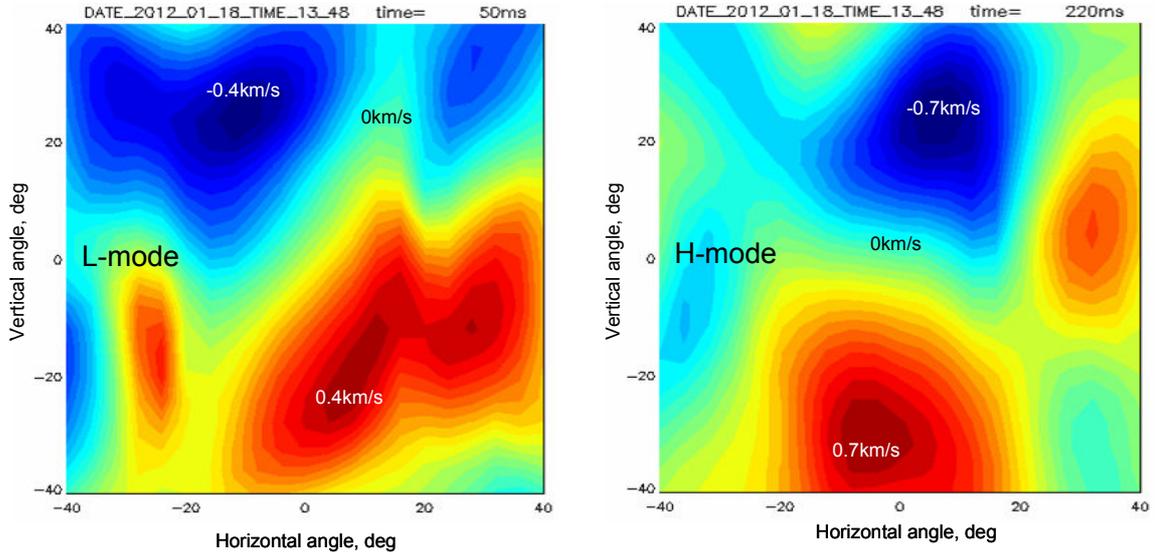

**Figure 11** Velocity maps of turbulence estimated from Doppler shifts of back-scattered signal at 10GHz. Note the flow directions are different during L-mode and H-mode phases.

Experimental results can be presented in several forms: 2D Doppler shift map for every frequency; a complementary 2D intensity map of the backscattered signal for every frequency; double-side band Fourier spectra for selected views and time intervals; weighted Doppler shift versus time for particular viewing direction and frequency. Fig. 11 shows velocity maps reconstructed from backscattered signals during so-called low-confinement and high-confinement mode operational phases. The differences in the flow direction and velocity values are clearly seen from comparison of these maps. Discussion of the physics explaining the difference in the flow directions is outside the scope of the present paper and will be published elsewhere. To our knowledge these are the first experimentally obtained 2D maps of the turbulence flow in fusion plasmas.

## 6. Conclusions

A microwave imaging system with ~10 μs time resolution has been designed and built for fusion plasma research. The system uses a phased array of antennas instead of focusing optics. The principle is based on the fact that phase differences between antenna pairs steer the viewing direction. Cross-correlations between pairs of antennas give the spatial Fourier transform of the emission pattern. The effective number of "pixels" is ~ $N^2$ (where $N$ is the number of antennas). A phased array of 8 antennas has been implemented on the MAST tokamak to record the first 2D images of thermal microwave emission escaping from the plasma.

The system records images at 16 discrete frequencies switching in a programmable manner within the range of 10-35 GHz (covering the range from $\omega_{ce}$ to $3 \cdot \omega_{ce}$). Frequency scanning provides radial resolution in addition to 2D images resulting in 2D+1f(R) resolution over the frequency scan. All signal processing stages including side band separation are performed digitally post-shot. Two images are generated at every frequency corresponding to the lower and upper side bands. Side band separation is about 0.1 GHz. First ever 2D images of mode converted Bernstein wave emission have been experimentally obtained from a fusion plasma. Experimental 2D images are in good agreement with mode coupling modelling results.

Simultaneously with passive imaging of the thermal emission from the plasma, the system performs active probing of the plasma surface. The probing signal is back-scattered by the plasma turbulence. The frequency of the scattered signal is Doppler shifted due to the turbulent motion. The system acquires scattered signals and produces a 2D velocity map of the plasma turbulence at every probing frequency. That is the first ever 2D turbulence velocity map experimentally measured in a fusion plasmas.

## Acknowledgments

The authors would like to thank colleagues A. Cross and K. Ronald from Strathclyde University for their help in the testing of our Vivaldi antennas. This work was funded partly by EPSRC under grant EP/H016732, by the University of York, by the RCUK Energy Programme under grant EP/I501045 and the European Communities under the contract of Association between EURATOM and CCFE. The views and opinions expressed herein do not necessarily reflect those of the European Commission.